\documentstyle[preprint,aps,tighten]{revtex}
\begin{document}
\draft
\title{
Comment on ``Relativistic cluster dynamics of nucleons and mesons. II.
Formalism
and examples''
}
\author{A.\ Stadler}
\address{
Department of Physics,
College of William and Mary, Williamsburg, Virginia 23185}
\author{J.\ A.\ Tjon}
\address{Institute for Theoretical Physics, University of Utrecht,
Princetonplein 5,
3508 TA Utrecht, The Netherlands}
\date{\today}
\maketitle

\begin{abstract}
In a recent paper [Phys.\ Rev.\ C{\bf 49}, 2142 (1994)], Haberzettl presented
cluster N-body equations for arbitrarily large systems of nucleons and mesons.
Application to the three-nucleon system is claimed to yield a new kind of
three-nucleon force. We demonstrate that these three-nucleon equations contain
double counting.
\end{abstract}

\narrowtext
Recently, H.\ Haberzettl proposed a relativistic formalism for the
nuclear N-body problem that is based on the
description of clusters rather than of individual particles \cite{Hab92,Hab94}.
In Ref.\ \cite{Hab94}, this formalism, which he calls
Relativistic Cluster Dynamics (RCD), is illustrated by several examples, among
them the
three-nucleon system with explicit pion and $\Delta$ degrees of freedom
in addition to the usual nucleonic ones. It is claimed that the RCD description
yields,
among others, a certain one-pion exchange three-nucleon force that has not been
included
in any three-nucleon calculation so far \cite{Hab94,Hab93}. In this comment, we
evaluate
the RCD graphs of
this one-pion exchange three-nucleon force using the ``RCD rules'' for the
calculation
of cluster graphs as they are specified in the Appendix of \cite{Hab94}. We may
compare
it to a graph that results from the iteration of the so-called nucleon-exchange
diagram
which describes three-nucleon scattering solely in terms of the two-nucleon
interaction.
We find that the contributions of the one-pion exchange three-nucleon force are
identical
to the part of the iterated nucleon-exchange diagram that is due to the pionic
component
of the two-nucleon interaction.
As a consequence, including the one-pion exchange three-nucleon force
amounts to double counting.

The basic equation of RCD is an effective two-body Lippman-Schwinger equation
of
the form
\begin{equation}
T^{\sigma \rho} = U^{\sigma \rho} + \sum_{\tau} U^{\sigma \tau} G_{0}^{\tau}
T^{\tau \rho} \, ,
\label{EqLS}
\end{equation}
where $T$ is the total scattering amplitude for the transition from a
two-cluster partition
$\rho$ to a two-cluster partition $\sigma$; $G_{0}^{\tau}$ describes the free
propagation of two
clusters, corresponding to the partition $\tau$, that do not interact with each
other but are
internally fully interacting. The driving term $U$ contains all diagrams that
are irreducible
with respect
to  $G_{0}^{\tau}$. Figure \ref{EqLS} displays Eq.\ (\ref{EqLS}) in graphical
form together with
the three diagrams of $U$ that we will focus on.

To demonstrate most clearly the double counting in these equations, let
us consider the lowest order connected field-theoretical
Feynman graph shown in Fig.\ 2. As a Feynman diagram, it sums all possible time
orderings of
pion exchanges, in particular also all processes where at least two pions are
exchanged in
overlapping time intervals.
Within the RCD approach two classes of contributions can be identified
with this Feynman graph. The diagrams $B_1$ and $B_2$ in the driving
force in Fig.\ 1 belong to one type, whereas there is also a contribution
from iterating the Lippmann-Schwinger equation once.

Let us first consider the class of diagrams as occurring in the
driving force in Fig.\ 1. In particular, we confine ourself to the
graphs $B_1$ and $B_2$, which have been argued in Refs.\ \cite{Hab94,Hab93}
as leading to the new three-nucleon force contributions.
If diagram $B_1$ of Fig.\ 1 is evaluated following the RCD
rules of Ref.\ \cite{Hab94},
we get (ignoring the three-momentum dependence and phase-space factors)
\widetext
\begin{eqnarray}
B_1  & \propto &
\frac{i}{2\pi}
\int dk_0 \, \left[g_{f} (k_0+e-e', E-e')\right]^\dagger
t_{\beta'}(k_0+e-e'-E_{\beta'}+i0)
\nonumber \\
& & \times
\left[f_{\delta\beta,\beta'}(e-e',k_0+e-e')\right]^\dagger
t_{\beta}(k_0-E_\beta +i0)
t_{\alpha}(E-e-k_0-E_\alpha +i0)
\nonumber \\
& & \times
\tau_{\delta}(e-e'-\omega_\delta +i0)
f_{\delta \gamma',\gamma}(e-e',e)
g_{i}(k_0,E-e)
\, ,
\label{EqB1}
\end{eqnarray}
\narrowtext
\noindent
where $t$ and $\tau$ are the nucleon and pion RCD propagators, $g$ and $f$ the
two-nucleon
and pion-nucleon
vertex functions (or ``form factors'' in the terminology of separable
interactions), and
$E_\alpha$ and $\omega_\delta$ are the on-shell energies of the nucleon
$\alpha$
and the pion
$\delta$, respectively.
Similarly, diagram $B_2$ yields
\widetext
\begin{eqnarray}
B_2  & \propto &
\frac{i}{2\pi}
\int dk_0 \, \left[g_{f} (k_0+e-e', E-e')\right]^\dagger
t_{\beta'}(k_0+e-e'-E_{\beta'}+i0)
\left[f_{\delta\gamma,\gamma'}(e'-e,e')\right]^\dagger
\nonumber \\
& & \times
t_{\beta}(k_0-E_\beta +i0)
t_{\alpha}(E-e-k_0-E_\alpha +i0)
\tau_{\delta}(e'-e-\omega_\delta +i0)
\nonumber \\
& & \times
f_{\delta \beta',\beta}(e'-e,k_0)
g_{i}(k_0,E-e)
\, .
\label{EqB2}
\end{eqnarray}
\narrowtext
\noindent
The precise meaning of the occurring energy variables and cluster indices is
defined in Fig.\ 3. Since energy (and momentum) is conserved at each vertex, it
is sufficient to label the form factors by two arguments only, the choice of
which
is just a matter of taste. We choose our notation such that it is easy to
follow
the ``flow of energy'' through the diagrams along the path that includes the
pion line. While the $\pi NN$ vertices carry subscripts for all connecting
clusters,
we characterize the $NN$ form factors only as initial ($i$) and final ($f$).
Assuming
point-like pion-nucleon vertices, we get
\widetext
\begin{eqnarray}
B_1 + B_2 & \propto &
\frac{i}{2\pi}
\Delta \, | f |^2
\int dk_0 \, \left[g_{f} (k_0+e-e', E-e')\right]^\dagger
t_{\beta'}(k_0+e-e'-E_{\beta'}+i0)
\nonumber \\
& & \times
t_{\beta}(k_0-E_\beta +i0)
t_{\alpha}(E-e-k_0-E_\alpha +i0)
g_{i}(k_0,E-e)
\, ,
\label{EqB3}
\end{eqnarray}
\narrowtext
\noindent
where $\Delta$ is the pion propagator,
\begin{equation}
\Delta = \tau_{\delta}(e-e'-\omega_\delta +i0) +
\tau_{\delta}(e'-e-\omega_\delta +i0) \, .
\end{equation}

In order to establish the relation with the Feynman graph of Fig.\ 2, we
note
that the two-nucleon form factors
$g_{i}$ and $g_{f}$ are directly related to the one-meson exchange
graphs through the separable expansion of the two-nucleon t-matrix.
In view of this, Eq.\ (4) can readily be identified as corresponding to
the Feynman graph of Fig.\ 2, with only the positive energy
states kept in the nucleon propagators and point-like pion-nucleon
vertices taken.

We now turn to the other class of diagrams as obtained from iterating the
integral equation.
If we iterate diagram $A$ of Fig.\ 1 once, we get the diagram depicted in
Fig.\ 4. Part of the iterated graph is a full two-nucleon t-matrix that
can be determined at the hierarchically lower two-body level. It is an
essential
feature
of the RCD
strategy to build up cluster amplitudes at the N body level from those of the
N-1 body level in a consistent fashion. The kernel of the two-nucleon cluster
equations from which the two-nucleon t-matrix is calculated is shown in Fig.\
17
of
Ref.\ \cite{Hab94}.
At this point we are only interested in the one-pion exchange
contributions to the two-nucleon interaction. The two-nucleon t-matrix, as
illustrated
in Fig.\ 5, consists of single pion-exchange processes plus all iterations
(for this discussion, any other contributions than those of pions are
irrelevant
and
not considered for simplicity).

If we keep only the lowest order pion-exchange contribution to the two-nucleon
t-matrix, the iterated one-nucleon exchange contribution reduces to
the graphs of Fig.\ 6.
We may note that the only difference between the two diagrams of  Fig.\ 6 and
diagrams
$B_1$ and $B_2$ of Fig.\ 1  consists
in a different arrangement of the pion production and absorption vertices,
respectively.
At first sight, this suggests that different physical processes are involved.
However,
the cluster diagrams of RCD should not be confused with diagrams of ordinary
time-ordered
perturbation theory, nor with Feynman diagrams. They are ``time''-ordered in
the
sense that
vertices can ``open'' or ``close'' as one follows the lines representing
clusters in a
given diagram (from right to left, according to the convention of Ref.\
\cite{Hab94}).
In other terms, a cluster can break up into two new clusters, and two clusters
can combine to a single one, respectively. Since the two processes are not
equivalent, the vertices
induce an ordering that resembles a ``time'' ordering. This is different in a
Feynman diagram where
a vertex describes absorption and emission simultaneously.
The rules of RCD as given in the Appendix of Ref.\ \cite{Hab94} assign vertex
functions
and propagators to each diagram; their energy and momentum dependence is
completely
determined by energy and momentum conservation at each vertex. It is therefore
irrelevant
whether, e.g., the pion production vertex of diagram $B_1$ is drawn on the
right
or left
side of the two-nucleon form factor $g_i$. Hence, the expressions for the
graphs
of
Fig.\ 6 are exactly the same as the ones for $B_1$ and $B_2$ of Fig.\ 1, namely
Eqs.\ (\ref{EqB1}) -- (\ref{EqB3}).
Of course, since we are dealing with the same expressions as for  $B_1$ and
$B_2$, it
follows immediately that identifying
the two-nucleon vertex functions through Fig.\ 5 as corresponding to the
one-pion exchange between the two nucleons leads again
to the Feynman graph of Fig.\ 2
with only the positive energy parts of the nucleon
propagators kept.

Hence, the Feynman graph
contribution of Fig.\ 2 is counted twice in the RCD equations.
Unlike in a nonrelativistic theory, iterating the two-nucleon
interaction automatically includes all time orderings of particle exchanges,
and
time-overlapping processes need not be added by hand.
{}From the above considerations it is obvious, that
the RCD equations contain in general double counting of a certain class of
fieldtheoretical graphs.
The basic reason for this double counting is that the structure
of the vertex functions is in principle dependent on the
underlying interaction. As a result, certain contributions from the
fieldtheoretical graphs will appear at different places in the RCD
approach, leading to double counting of graphs.

Support by the U.S.\ Department of
Energy under grant DE-FG05-88ER40435 (A.S.) is gratefully acknowledged.
J.A.T.\ thanks the nuclear theory group at CEBAF for its
hospitality during the visits when this work was done.

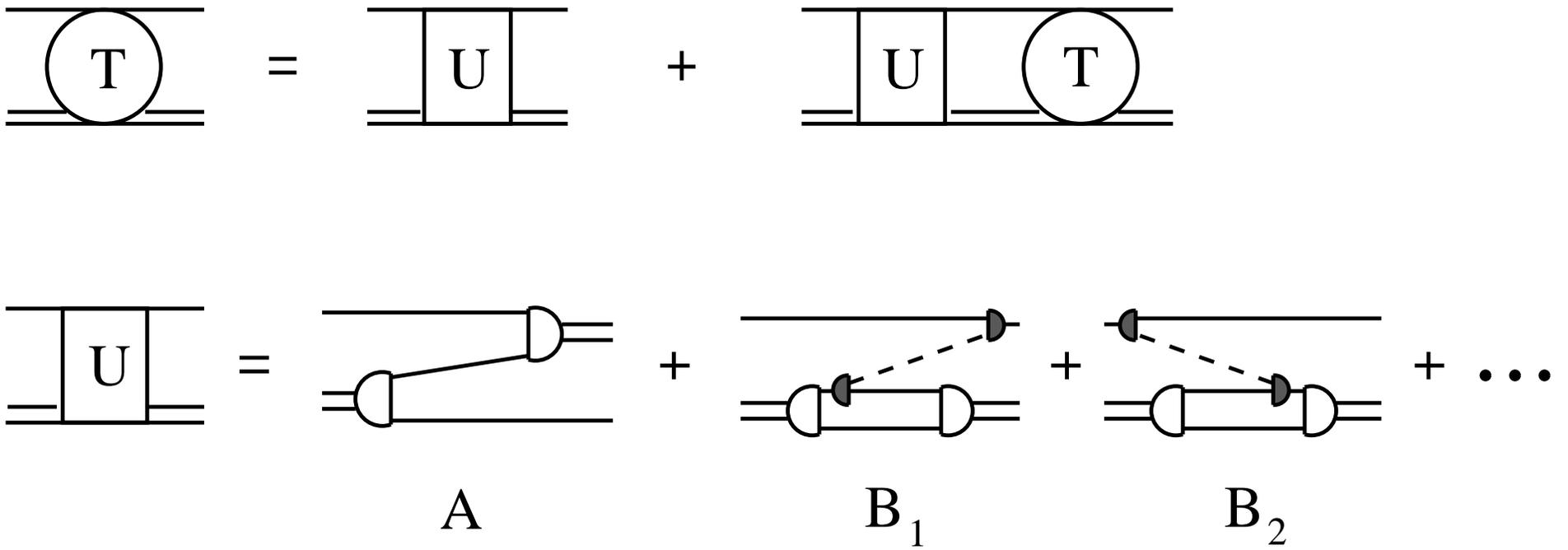
\begin{figure}
\caption{Lippman-Schwinger type equation for three-body scattering. In all
figures, solid lines
are nucleons, dashed lines are pions. Open
semicircles are two-nucleon form factors, hashed semicircles are pion
production and absorption vertices.
}
\end{figure}

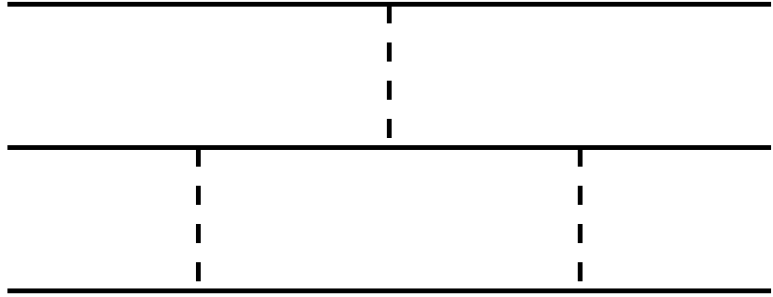
\begin{figure}
\caption{Lowest order connected pion exchange Feynman graph contribution
to the three-nucleon t-matrix.}
\end{figure}

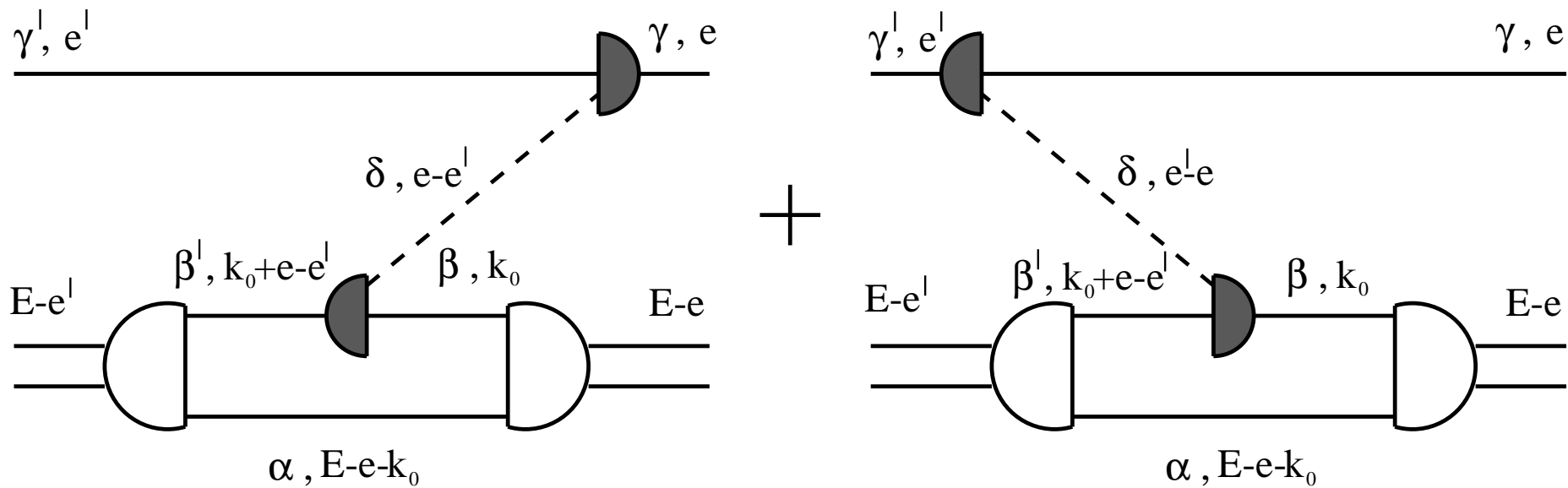
\begin{figure}
\caption{Energy and cluster variables for the evaluation of diagrams $B_1$ and
$B_2$ of
Fig.\ 1}
\end{figure}

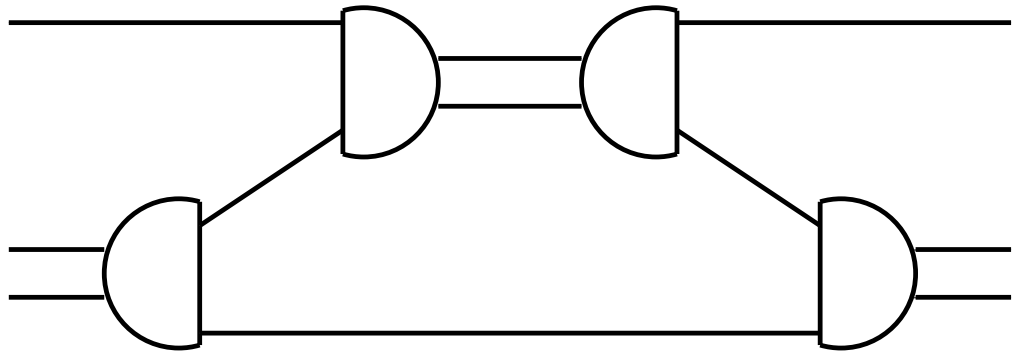
\begin{figure}
\caption{Iterated one-nucleon exchange graph $A$ of Fig.\ 1.}
\end{figure}

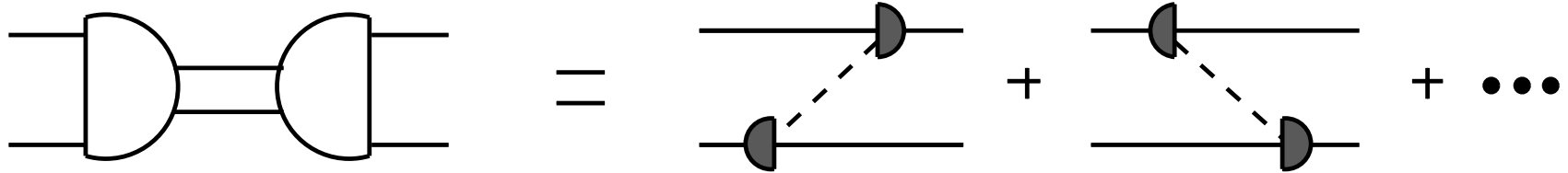
\begin{figure}
\caption{Lowest order pionic contributions to the two-nucleon t-matrix.}
\end{figure}

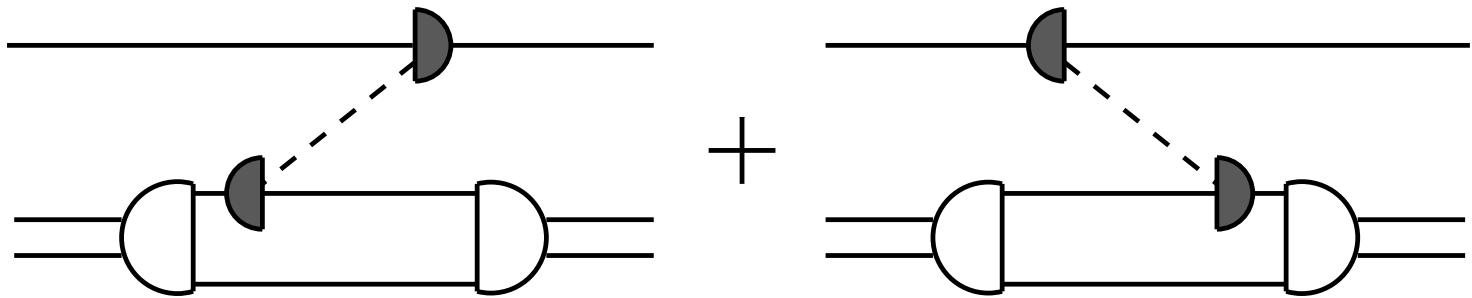
\begin{figure}
\caption{Lowest order pionic contributions to the iterated one-nucleon exchange
term of Fig.\ 4.}
\end{figure}
\end{document}